\documentclass[aps,showpacs,amsmath,amssymb,nofootinbib,preprintnumbers]{revtex4}
\usepackage{graphicx}
\usepackage{color}

\def \be  {\begin{equation}}
\def \ee  {\end{equation}}
\def \bea {\begin{eqnarray}}
\def \eea {\end{eqnarray}}

\begin{document}

\preprint{ECTP-2010-11}

\title{Antiproton-to-Proton Ratios for ALICE Heavy-Ion Collisions}
\author{A.~Tawfik}
\email{drtawfik@mti.edu.eg}
\affiliation{Egyptian Center for Theoretical Physics (ECTP), MTI University,
Cairo-Egypt}

\date{\today}

\begin{abstract}
Assuming that the final state of hadronization takes place along the freezeout line, which is defined by a constant entropy density, the antiproton-to-proton ratios produced in heavy-ion collisions are studied in framework of the hadron resonance gas (HRG) model. A phase transition from quark--gluon plasma to hadrons, a hadronization, has been conjectured in order to allow modifications in the phase space volume and thus in single--particle distribution function. Implementing both modifications in the grand--canonical partition function and taking into account the experimental acceptance in heavy-ion collisions, the antiproton-to-proton ratios over center-of-mass energies $\sqrt{s}$ ranging from AGS to RHIC are very well reproduced by the HRG model. Comparing with the same particle ratios in $pp$ collisions results in a gradually narrowing discrepancy with increasing $\sqrt{s}$. At LHC energy, the ALICE antiproton-to-proton ratios in $pp$ collisions turn to be very well described by HRG model as well. It is likely that the ALICE heavy-ion program will produce the same antiproton-to-proton ratios as the $pp$ program. Furthermore, the ratio gets very close to unity indicating that the matter-antimatter asymmetry nearly vanishes. The chemical potential calculated at this energy strengthens the assumption of almost fully matter-antimatter symmetry at LHC energy. 
\end{abstract}



\maketitle

\section{Introduction}

The collective properties of hot and dense QCD matter is one of the main objectives of the heavy-ion program. The possible modification of the transport coefficients, like phase structure and effective degrees of freedom, with increasing incident energy and system size likely provides fruitful tools to study the collective properties. Furthermore, it comprehensively characterizes the particle production itself \cite{jeon}.
The universal description for the particle production at different center-of-mass energies of nucleon-nucleon collision $\sqrt{s}$ can be achieved by studying the fluctuations \cite{rajan,jeon} and multiplicities of particle ratios~\cite{Sollfrank:1999cy,Cleymans:1998yf}, which are not only able to determine the freezeout parameters, temperature $T$ and chemical potential $\mu$, but also - to a large extend - eliminate the volume fluctuations and the dependence of the freezeout surface on the initial conditions. 
The fluctuations in the particle production are also essential regarding with examining the existing statistical models \cite{giorg,Karsch:2003vd}, characterizing the particle equilibration \cite{pequil} and search for unambiguous signals for the creation of the quark-gluon plasma (QGP) \cite{qgpA,qgpB}. 

Combining various fluctuations and multiplicities of integrated particle ratios~\cite{Sollfrank:1999cy,Cleymans:1998yf} makes it possible to determine the freezeout parameters by means of thermal models. The search for common properties of the freezeout parameters in heavy-ion collisions at different center-of-mass energies has a long tradition~\cite{Cleymans:1999st,Braun-Munzinger:1996mq}. Different models have been suggested~\cite{Cleymans:1999st,Braun-Munzinger:1996mq,Magas:2003wi,Tawfik:2005qn,Tawfik:2004ss}. In present work, the constant entropy density $s$ normalized to $T^3$ \cite{Tawfik:2004ss} is used. It implies that the hadronic matter in rest frame of produced system is characterized by constant degrees of freedom. The scaling of the quantity $s/T^3$ by $\pi^2/4$, which reflects the phase space volume, obviously results in the effective degrees of freedom of free gas~\cite{Cleymans:2005km}. 

On the other hand, it has been concluded that none of statistical models \cite{statst} would be able to perfectly describe the experimental data. Using grand--canonical ensembles and phase space modification factor ${\cal Q}$ leads to a much better description of the experimental data \cite{Tawfik:2010uh,Tawfik:2010aq}. The hadron resonance gas (HRG) model turns to fairly reproduce the particle multiplicities, the dynamical fluctuations of various particle ratios \cite{tawPS} and the thermodynamics of the strongly interacting system. In doing this, ${\cal Q}$ plays an essential role. It has been noticed that the fluctuations over the whole range of $\sqrt{s}$ exhibit a non-monotonic behavior and ${\cal Q}$ is varying with increasing $\sqrt{s}$ \cite{Tawfik:2010uh,Tawfik:2010aq}.

Enhanced antibaryons are conjectured as indicators for the formation of deconfined QGP \cite{antib1}, whereas the possible annihilation might suppress such enhancement \cite{antib2}. The collision initial conditions and formation time can be reflected by surviving antibaryons, where antiproton is a good candidate. Decelerating or even stopping of incident particle and its break up in inelastic collisions have been discussed in literature \cite{pbreak}. Therefore, $n_{\bar{p}}/n_p$ ratios can be used to study the baryon transport and production and therefore would have significant astrophysical consequences. By mentioning astrophysics, it seems in order to recall now that the $n_{\bar{p}}/n_p$ ratios have been calculated and also observed in the cosmic rays revealing essential details on astrophysical phenomena \cite{antip_cr}. In present work, $n_{\bar{p}}/n_p$ ratios are calculated in HRG model and compared with the heavy-ion collisions at $\sqrt{s}$ ranging from AGS to LHC. Thus, $n_{\bar{p}}/n_p$ ratios and their description in the thermal model seem to provide fruitful tools to study the evolution of matter-antimatter asymmetry with changing energy. In section \ref{sec:2}, the model is introduced. Section \ref{sec:rslts} is devoted to the results and conclusions.

\section{Hadron Resonance Gas Model}
\label{sec:2}

The hadron resonances treated as a free
gas~\cite{Karsch:2003vd,Karsch:2003zq,Redlich:2004gp,Tawfik:2004sw,Taw3} are
conjectured to add to the thermodynamic quantities in the hadronic phase of heavy-ion collisions. This
statement is valid for {\it free} as well as {\it strong} interactions between the
hadron resonances themselves. It has been shown that the thermodynamics of strongly interacting
system can be approximated to an ideal gas composed of hadron
resonances~\cite{Tawfik:2004sw,Taw3,Vunog}. Also, it has been shown that the thermodynamics of strongly interacting system can be approximated to an ideal gas composed of hadron resonances with masses $\le 1.8~$GeV~\cite{Tawfik:2004sw,Taw3,Vunog}. The heavier constituents, the smaller thermodynamic quantities. The main motivation of using the Hamiltonian is that it contains all relevant degrees of freedom of {\it confined} and {\it strongly} interacting matter. It implicitly includes the interactions that result in the formation of {\it new} resonances. In addition, HRG model is used to provide a quite satisfactory description of particle production and collective properties in heavy-ion collisions \cite{Tawfik:2005qn,Tawfik:2010uh,Tawfik:2010aq,tawPS,Redlich:2004gp,Karsch:2003zq,Tawfik:2004sw,Taw3,Vunog}. 

At finite temperature $T$, strange $\mu_S$ and isospin  $\mu_{I_3}$
and baryochemical potential $\mu_B$, the partition function of one single hadron reads \cite{Tawfik:2010uh,Tawfik:2010aq}
\bea
\label{eq:zTr2Tsl}
\ln Z_{gc}(T,V,\mu)&=& \sum_i\pm \frac{g_i}{2\pi^2}\,V\int_0^{\infty} k^2 dk\,\ln\,\left(1\pm \left[1+\left(\frac{\epsilon_i(\vec{k})-\mu_i}{T}-\alpha\right)(q-1)\right]^{1/(q-1)}\right), \\
\label{eq:zTr2}
\ln Z_{gc}(T,{\cal V},\mu)&=& \sum_i\pm \frac{g_i}{2\pi^2}\,{\cal V}\int_0^{\infty} k^2 dk\; \ln\,\left(1\pm \gamma\,{\cal Q}\, \exp\left[\frac{\mu_i-\epsilon_i(k)}{T}\right]\right),
\eea
where $\varepsilon_i(k)=(k^2+m_i^2)^{1/2}$ is the $i$-th particle's dispersion relation (single-particle energy) in equilibrium and $\pm$ stands for bosons and fermions, respectively. $g$ is the spin-isospin degeneracy factor and
$\gamma\equiv\gamma_q^n\gamma_s^m$ stand for the quark phase
space occupancy parameters, where $n$ and $m$ being number of light and
strange quarks, respectively. $\lambda=\exp(c\,\mu/T)$ is the
fugacity, where $\mu$ is the chemical potential multiplied by corresponding
charge $c$, like baryon, light and strange quark numbers, etc. Summing over all hadron resonances results in the final 
thermodynamics in the hadronic phase, since no phase transition is 
conjectured in HRG model. 

Parameters $\alpha$ and $\beta$, are Lagrange multipliers in the entropy maximization. The physical meaning of $\alpha$ is a controller over the number of particles in the phase space, i.e, acting as chemical potential, $\alpha=\ln{\cal E}-\ln T-\ln N$.
It also combines intensive variables, $T$ and $N$ with an extensive one ${\cal E}=\sum_i^n g_i\epsilon_i$, where $\epsilon_i$ is energy density of $i$-th cell in the phase space of interest. 
The most probable state density can be found by the Lagrange multipliers, where one of them, $\alpha$, has been expressed in term of the second one, $\beta$, and the occupation numbers of the system. Apparently, $\alpha$ gives how the energy ${\cal E}$ is distributed in the microstates of the equilibrium system and therefore, can be understood as another factor controlling the number of occupied states at microcanonical level.

The generic parameter ${\cal Q}(\vec{x},\vec{k})$ is conjectured to reflect the change in phase-space, when the hadronic degrees of freedom are replaced by partonic ones and vice versa. Also, it can be interpreted as a measure for the non-extensivity. Comments on the statistical parameter $\gamma=\exp(-\alpha)$ are now in order. 
It gives the averaged occupancy of the phase space relative to
equilibrium limit. Assuming finite time evolution of the system, $\gamma_i$ can be seen as a ratio of the change in particle number before and after the chemical freeze-out, i.e. $\gamma_i=n_i(t)/n_i(\infty)$. The chemical freezeout is defined as a time scale, at which there is no longer particle production and the collisions become entirely elastic. 
In case of phase transition, $\gamma_i$ is expected to
be larger than one, because of the large degrees of freedom,
weak coupling and expanding phase space above the critical temperature to QGP.

The quark chemistry is given by relating the {\it hadronic} chemical potentials and
$\gamma$ to the quark constituents. $\gamma\equiv\gamma_q^n\gamma_s^m$ with
$n$ and $m$ being the number of light and strange quarks,
respectively. $\mu_B=3\mu_q$ and $\mu_S=\mu_q-\mu_s$, with $q$ and $s$ are
the light and strange quark quantum number, respectively. The
baryochemical potential for the light quarks is
$\mu_q=(\mu_u+\mu_d)/2$. The iso-spin
chemical potential $\mu_{I_3}=(\mu_u-\mu_d)/2$. Therefore, vanishing $\mu_{I_3}$ obviously means that light quarks are conjectured to be degenerate and $\mu_q=\mu_u=\mu_d$. The strangeness chemical potential, $\mu_S$, is calculated in HRG model as a function of $T$ and $\mu_B$. At finite $T$ and $\mu_B$,  $\mu_S$ is calculated under the condition of the overall strangeness conservation. 

Absolving chemical freezeout means that the hadrons finally decay to stable hadrons or other resonances
\bea \label{eq:n2}
n_i^{final} &=& n_i^{direct} + \sum_{j\neq i} b_{j\rightarrow i} \; n_j,
\eea
where $b_{j\rightarrow i}$ being decay branching ratio of $j$-th hadron resonance
into $i$-th {\it stable} particle. 
The particle number density in grand-canonical ensemble is no longer constant. In an ensemble of $N_r$ hadron resonances,
\bea 
\label{eq:n1} 
n &=& \sum_i^{N_r} \frac{g_i}{2\pi^2} \int_0^{\infty} k^2 dk 
\frac{{\cal Q}\,\gamma}{\exp\left[\frac{\epsilon_i(k)-\mu_i}{T}\right] \pm {\cal Q}\,\gamma}.
\eea

In carrying out the calculations given in Fig. \ref{fig:1a}, a full grand-canonical statistical
set of the thermodynamic parameters has been used. Corrections due to van~der~Waals
repulsive interactions has not been taken into account~\cite{Tawfik:2004sw}.
Seeking for simplicity, one can for a moment assume the Boltzmann approximations, especially at ultra-relativistic energy, in order find out the parameters, on which $n_{\bar{p}}/n_p$ ratio depends. Then, at finite isospin fugacity $\lambda_{I_3}$,  
\bea
\frac{n_{\bar{p}}}{n_{p}}  &\simeq& 
\frac{\lambda_{\bar{p}}}{\lambda_{p}} 
      = \left(\lambda_{\bar{u}}^2\; \lambda_{\bar{d}}\right)^2. \label{naptip-p} 
\eea

\section{Results and Conclusions}
\label{sec:rslts}

As mentioned above, the formation of deconfined QGP strongly depends on the initial conditions in the early stage of heavy-ion collisions, where the baryon number transport (deceleration of the incoming particle) or even its entire stopping is assumed to be achieved \cite{nb1}. Therefore, it is widely conjectured that  such deconfined matter would not be produced, especially in $pp$ collisions at low energies. On the other hand, the degrees of baryon number transport likely affects the overall dynamical evolution of the system. The initial parton equilibrium \cite{ipd}, thermal/chemical equilibrium \cite{tce}, time evolution  \cite{tev} and the final particle production \cite{pp} are affected as well. 

\begin{figure}[thb]
\includegraphics[width=8.cm,angle=-90]{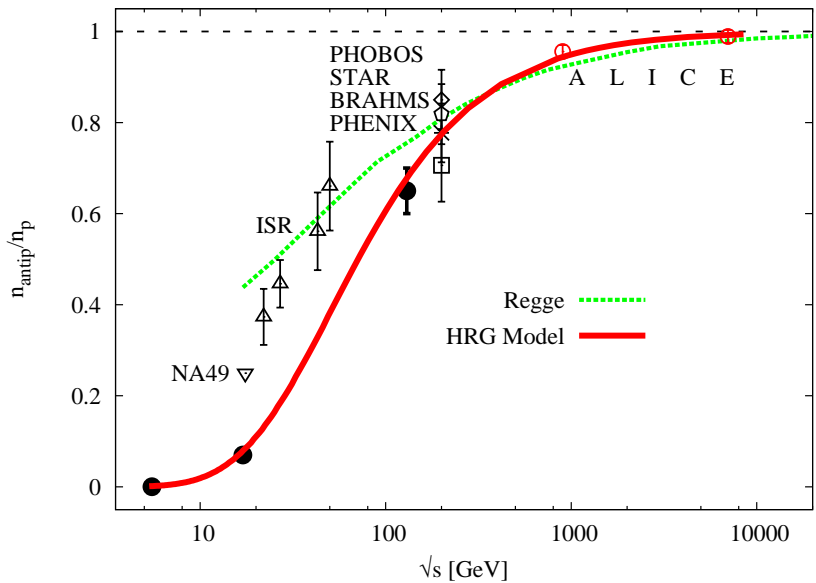}
\caption{\normalsize $n_{\bar{p}}/n_p$ ratios depicted in the whole available range of $\sqrt{s}$. Open symbols stand for the results from various $pp$ experiments (labeled). The solid symbols give the heavy-ion results from AGS, SPS and RHIC, respectively. The fitting of $pp$ results according to Regge model is given by the dashed curve \cite{alice2010}. The solid curve is the HRG results at ${\cal Q}=0.5$ and $\gamma=1.0$, which apparently perfectly describes all heavy-ion collisions, besides the ALICE $pp$ results. Extending it to the LHC energy obviously shows that the ratio itself is very much close to unity. Contrary to the dashed curve, the solid line is not a fitting to experimental data.} 
\label{fig:1a}
\end{figure}

In Fig. \ref{fig:1a}, $n_{\bar{p}}/n_p$ ratios are given in dependence on center-of-mass-energy $\sqrt{s}$ ranging from AGS to LHC. The $pp$ collisions (empty symbols) are compared with the heavy-ion collisions (solid symbols). In the nucleus-nucleus collisions, the spatial and time evolution of the system is too short to assure initial conditions required in order to derive the parton matter into deconfined QGP. We notice that $n_{\bar{p}}/n_p$ ratios up to RHIC energy are smaller than unity indicating an overall excess of $p$ over $\bar{p}$. This would also imply that a certain fraction of the baryon number is likely transported from the incident projectile at beam rapidity to mid-rapidity region of the system. Thus up to RHIC energy, the mid-rapidity region seems to have finite net-baryon number. 

As mentioned above, HRG model has been successfully utilized to describe heavy-ion collisions and the thermodynamics of lattice QCD as well \cite{Karsch:2003vd,Tawfik:2005qn,Tawfik:2004ss,Karsch:2003zq,Redlich:2004gp,Tawfik:2004sw}. The collective flow of the matter in heavy-ion collisions makes it obvious that HRG model underestimates the $n_{\bar{p}}/n_p$ ratios measured in the $pp$ collisions, especially at low energies. On the other hand, the ratios from the heavy-ion collisions \cite{star1} are very well reproduced by HRG model. In both types of collisions, there is a dramatic increase in the $n_{\bar{p}}/n_p$ ratio. For instance, in RHIC heavy-ion collisions, the ratio is $0.65\pm 0.05$ ($\sqrt{s}=130\,$GeV), which is obviously much higher than the ones measured at  SPS ($\sim0.07\pm10\%$ at $\sqrt{s}=17\,$GeV) \cite{SPS1} and AGS ($\sim2.5\times10^{-4}\pm10\%$ at $\sqrt{s}=5\,$GeV) \cite{AGS1}. Also in the $pp$ collisions, the $n_{\bar{p}}/n_p$ ratios obviously increase from NA49 over ISR and $200\,$GeV-RHIC to ALICE. Up to RHIC energy, it is obvious that the matter-antimatter asymmetry decreases from nearly $100\%$ at AGS to $65\%$ and $130\,$GeV-RHIC. At LHC energy, while the differences between $n_{\bar{p}}/n_p$ ratios of $pp$ and heavy-ion collisions apparently disappear, their values approach unity, i.e. almost $0\%$ matter-antimatter asymmetry.  

The dashed line in Fig. \ref{fig:1a} gives the fitting according to Regge model, which reasonably describe the $n_{\bar{p}}/n_p$ data from different $pp$ collisions. This fitting is elaborated in Ref. \cite{alice2010}. The baryon pair production is assumed to be governed by pomeron exchange and baryon transport by string-junction exchange \cite{reggee}. Therefore, it has been concluded that ALICE results seem to be consistent with the baryon number transport.

The solid line gives the results from HRG model, section \ref{sec:2}. No fitting has been utilized. Over the whole range of $\sqrt{s}$, the two parameters $\gamma$ and ${\cal Q}$ remain constant, $1.0$ and $0.5$, respectively. It seems to be a kind of a universal curve. It describes very well the heavy-ion results. Also ALICE $pp$ results are reproduced by means of HRG model. So far, it can be concluded that the future heavy-ion program of ALICE likely will produce $n_{\bar{p}}/n_p$ ratios similar, when not entirely identical, to the ones measured in this current $pp$ program. Secondly, the ratios run very close to unity implying almost vanishing matter-antimatter asymmetry. Some astrophysical consequences of these results will shortly be elaborated below. On the other hand, it can also be concluded that the thermal models including HRG seem to be able to perfectly describe the hadronization at very large energies and the condition deriving the chemical freezeout at the final state of hadronization, the constant degrees of freedom or $S(\sqrt{s},T)=7 (4/\pi^2) V T^3$, seems to be valid at all center-of-mass-energies raging from AGS to LHC.  

Expression (\ref{naptip-p}) can be used to calculate the antiproton (antiquark) chemical potential. At a vanishing isospin chemical potential $\mu_{I_3}$, i. e. degenerate light quarks 
\bea
\frac{n_{\bar{p}}}{n_{p}}  &\simeq&  \exp(-2\mu_{\bar{p}}/T_{ch}) 
      = \exp(-6\mu_{\bar{q}}/T_{ch}), \label{naptip-p2} 
\eea
where $T_{ch}$ is the freezeout temperature. Substituting $n_{\bar{p}}/n_p$ of ALICE experiment at $7\,$GeV and $T_{ch}=0.175\pm0.02\,$GeV in Eq. (\ref{naptip-p2}) results is
\bea
\mu_{\bar{p}}=3\mu_{\bar{q}} &\simeq& 0.537\;\;\mathtt{MeV} \label{naptip-p3} 
\eea
The ratio of baryon density asymmetry  to photon density, $\eta$, has been measured in WMAP data \cite{wmap}, then $n_{\bar{p}}-n_p$ from ALICE experiment at $7\,$GeV gives a photon number density ranging between
\bea
4.4\times 10^{4}\; < & n_{\gamma} & <\; 7.7\times 10^{4}.
\eea
When modelling the volume of the early universe at energies equal to top ALICE, $7\,$TeV, the photon number, $N_{\gamma}=n_{\gamma}\, V$, can be calculated. It is obvious that the resulting $N_{\gamma}$ will be extraordinarily large indicating dominant radiation against all other equations of state in this era where matter and antimatter are supposed to be almost symmetric.

The matter-antimatter asymmetry is one of the greatest mysteries in modern physics. The well-known Sakharov's conditions \cite{sakhrv}; baryon number violation and C and CP violation besides out-of-equilibrium processes; are conjectured to give a solid framework to explain why should matter become dominant against antimatter with the universe cooling-down of expansion. Recently, a fourth condition has been suggested; $b$-$l$ violation, where $b$ and $l$ being baryon and lepton quantum number, respectively. In standard model, both $b$ and $l$ are not conserved, explicitly. Non-perturbatively, they can be violated through sphalerons, for instance, which derive the two-direction conversion $b\leftrightarrow l$ with a certain number of selection rules \cite{sphl}. That the heavy-ion program seems to give systematic tools to survey the evolution of matter-antimatter asymmetry from $\sim 0\%$ to $\sim 100\%$ in the nuclear interactions would make it possible to devote such experimental facilities to study the evolution of Sakharov's conditions.


\begin{thebibliography}{99}

\bibitem{jeon} S. Jeon and V. Koch, Phys. Rev. Lett. {\bf 83}, 5435 (1999).

\bibitem{rajan} M. Stephanov, K. Rajagopal and E. Shuryak, Phys. Rev. D {\bf 60}, 114028 (1999).

\bibitem{Sollfrank:1999cy}
J.~Sollfrank, U.~W. Heinz, H.~Sorge, and N.~Xu.
J. Phys. G {\bf 25}, 363 (1999).

\bibitem{Cleymans:1998yf}
H.~Cleymans, J.~Oeschler and K.~Redlich.
J. Phys. G {\bf 25}, 281--285 (1999).

\bibitem{giorg} G.~Torrieri, S.~Jeon and J.~Rafelski, Phys. Rev. C {\bf 74}, 024901 (2006).

\bibitem{Karsch:2003vd}
F.~Karsch, K.~Redlich, and A.~Tawfik.
Eur. Phys. J. C {\bf 29}, 549--556 (2003).

\bibitem{pequil} Q. H. Zhang, V. Topor Pop, S. Jeon and C. Gale, Phys.
Rev. C {\bf 66}, 014909 (2002).

\bibitem{qgpA} A. Bialas and R. C. Hwa, Phys. Lett. B {\bf 253}, 436 (1991).

\bibitem{qgpB} S. Hegyi and T. Csorgo, Phys. Lett. B {\bf 296}, 256 (1992).

\bibitem{Cleymans:1999st}
J.~Cleymans and K.~Redlich.
Phys. Rev. C {\bf 60}, 054908 (1999).

\bibitem{Braun-Munzinger:1996mq}
P.~Braun-Munzinger and J.~Stachel.
Nucl. Phys. A {\bf 606}, 320--328 (1996).

\bibitem{Tawfik:2004ss} A.~Tawfik, Europhys. Lett. {\bf 75}, 420 (2006). 

\bibitem{Magas:2003wi} V.~Magas and H.~Satz,
Eur. Phys. J.  C {\bf 32}, 115--119 (2003).

\bibitem{Tawfik:2005qn} A.~Tawfik, Nucl. Phys. A {\bf 764}, 387-392 (2006).

\bibitem{Cleymans:2005km}
J.~Cleymans, M.~Stankiewicz, P.~Steinberg, and S.~Wheaton,
nucl-th/0506027

\bibitem{statst} G. Torrieri, R. Bellwied, C. Markert and G. Westfall, int. conf. SQM 2009, Buzios, Brazil, 27 Sep - 2 Oct 2009. 

\bibitem{Tawfik:2010uh}
     A. Tawfik,
     arXiv:1007.4074 [hep-ph] 

\bibitem{Tawfik:2010aq} 
     A. Tawfik,
     arXiv:1007.4585 [hep-ph]  

\bibitem{tawPS} A. Tawfik, Fizika B {\bf 18}, 141-150 (2009); 0805.3612 [hep-ph]; hep-ph/0602094.

\bibitem{antib1} U. Heinz, P. R. Subramanian, W. Greiner, Z. Phys A {\bf 318}, 247 (1984); P. Koch, B. Muller, H. Stocker, W. Greiner, Mod. Phys. Lett. A {\bf 3}, 737 (1988); J. Ellis, U. Heinz, H. Kowalski, Phys. Lett. B {\bf 233}, 223 (1989).

\bibitem{antib2} S. Gavin, M. Gyulassy, M. Plumer, R. Venugopalan, Phys. Lett. B {\bf 234}, 175 (1990); H. Sorge, A. V. Keitz, R. Mattiello, H. Stocker, W. Greiner, Z. Phys. C {\bf 47}, 629 (1990).

\bibitem{pbreak} G.C. Rossi and G. Veneziano, Nucl. Phys. B {\bf 123}, 507 (1977); A. Capella {\it et al.} Phys. Rep. {\bf 236}, 225 (1994); A. B. Kaidalov and K. A. Ter-Martirosyan, Sov. J. Nucl. 411 Phys. {\bf 39}, 1545 (1984).

\bibitem{antip_cr} A. W. Labrador and R. A. Mewaldt, Astrophys. J. {\bf 480}, 480 (1997); S.P. Ahlen {\it et al.}, Phys. Rev. Lett. {\bf 61}, 145-148 (1988); T. K. Gaisser and B. G. Mauger, Astrophys. J. {\bf 252}, L57-L59 (1982). 

\bibitem{Karsch:2003zq}
F.~Karsch, K.~Redlich, and A.~Tawfik.
Phys. Lett. B {\bf 571}, 67--74 (2003).

\bibitem{Redlich:2004gp}
K.~Redlich, F.~Karsch, and A.~Tawfik.
J. Phys. G {\bf 30}, S1271--S1274 (2004).

\bibitem{Tawfik:2004sw}
A.~Tawfik.
Phys. Rev. D {\bf 71}, 054502 (2005).

\bibitem{Taw3}A.~Tawfik, J.~Phys.~G~{\bf 31}, S1105 (2005).


\bibitem{Vunog}R.~Venugopalan, M.~Prakash, Nucl.~Phys.~A~{\bf 546},~718~(1992).


\bibitem{nb1} W. Busza and A.S. Goldhaber, Phys. Lett. B {\bf 139}, 235 (1984);
        S. E. Vance, M. Gyulassy, and X. N. Wang, Phys. Lett. B {\bf 443}, 45 (1998); S. E.    Vance,  Nucl. Phys. A {\bf 661}, 230c (1999).

\bibitem{ipd} M. Gyulassy and X.N. Wang, Nucl. Phys. B {\bf 420}, 583 (1994); 
              R. Baier, Yu. L. Dokshitzer, S. Peigne, and D. Schiff, Phys. Lett. B {\bf 345}, 277 (1995).

\bibitem{tce} P. Braun-Munzinger, I. Heppe, and J. Stachel, Phys. Lett. B {\bf 465}, 15 (1999).

\bibitem{tev} I.G. Bearden {\it et al.}, (NA44 Collaboration), Phys. Rev. Lett. {\bf 78}, 2080 (1997); K.H. Ackermann {\it et al.}, (STAR Collaboration), Phys. Rev. Lett., {\bf 86}, 402 (2001).

\bibitem{pp} K. H. Ackermann {\it et al.} [STAR Collaboration], Nucl. Phys. A {\bf 661} 681-685 (1999).

\bibitem{star1} C. Adler {\it et al.}, Phys. Rev. Lett. B {\bf 86}, 4778 (2001).

\bibitem{SPS1} F. Sickler {\it et al.}, [NA49 collaboration], Nucl. Phys. A {\bf 661}, 45c (1999). 

\bibitem{AGS1} L. Ahle {\it et al.}, The E802 collaboration, Phys. Rev. Lett. {\bf 81}, 2650 (1998).

\bibitem{alice2010} K. Aamodt {\it et al.}, [Alice` Collaboration], Phys. Rev. Lett {\bf 105}, 072002 (2010).

\bibitem{reggee} D. Kharzeev, Phys. Lett. B {\bf 378}, 238 (1996).

\bibitem{wmap} C. Bennett {\it et al.}, [WAMP Collaboration], App. J. Suppl. {\bf 148}, 1 (2003); A. Tawfik, AIP Conf. Proc. {\bf 1115}, 239-247 (2009).

\bibitem{sakhrv} A. D. Sakharov, JETP Lett. {\bf 5}, 24 (1967).

\bibitem{sphl} V. A. Kuzmin, V. A. Rubakov and M. E. Shaposhnikov, Phys. Lett. {\bf 155}, 36 (1985); A. Riotto and M. Trodden, Ann. Rev, Part. Nucl. Sci. {\bf 49}, 35 (1999).


\end{thebibliography}
\end{document}